\documentclass{elsart}

\usepackage{amssymb}
\usepackage{graphicx}

\def\BibTeX{{\rm B\kern-.05em{\sc i\kern-.025em b}\kern-.08em
    T\kern-.1667em\lower.7ex\hbox{E}\kern-.125emX}}

\begin{document}

\begin{frontmatter}

\title{Quantifying information loss on chaotic attractors through recurrence networks}

\author[label1]{K. P. Harikrishnan\corauthref{cor1}}
\ead{$kp.hk05@gmail.com$}
\author[label2]{R. Misra}
\ead{$rmisra@iucaa.in$}
\author[label3]{G. Ambika}
\ead{$g.ambika@iisertirupati.ac.in$}

\corauth[cor1]{Corresponding author: Address: Department of Physics, The Cochin College,  Cochin-682002, India; Phone No.0484-22224954;  Fax No: 91-22224954.} 

\address[label1]{Department of Physics, The Cochin College, Cochin-682002, India}
\address[label2]{Inter University Centre for Astronomy and Astrophysics, Pune-411007, India}
\address[label3]{Indian Institute of Science Education and Research, Tirupati-517507, India}

\begin{abstract}
We propose an entropy measure for the analysis of chaotic attractors through recurrence networks 
which are un-weighted and un-directed complex networks constructed from time series of dynamical 
systems using specific criteria. We show that the proposed measure converges to a constant 
value with increase in the number of data points on the attractor (or the number of nodes on the 
network) and the embedding dimension used for the construction of the network, and clearly 
distinguishes between the recurrence network from chaotic time series and white noise. Since the 
measure is characteristic to the network topology, it can be used to quantify the information loss 
associated with the structural change of a chaotic attractor in terms of the difference in the link 
density of the corresponding recurrence networks. We also indicate some practical applications of the 
proposed measure in the recurrence analysis of chaotic attractors as well as the relevance of the proposed 
measure in the context of the general theory of complex networks.  
\end{abstract}


\end{frontmatter}

\section{Introduction}
Chaotic dynamical systems are those whose evolution causes exponential divergence of nearby 
trajectories \cite {spr}. This, in turn, leads to a loss of information during the evolution of 
the system which is quantified using an entropy measure called Kolmogorov-Sinai entropy \cite {hil} 
or K-S entropy. In the case of chaotic attractors generated from time series, a related measure 
called correlation entropy \cite {ken} denoted by $K_2$ is used for this purpose.  

In the case of a fully developed chaotic attractor, there can also be a loss of information as a 
result of a structural change on the attractor \cite {aba}, which is clearly different from the 
information loss arising out of the dynamical evolution of the system mentioned above. An obvious 
example is the effect of contamination by noise which changes the structure of the attractor.  
To quantify such an information loss, we require tools that can capture the structural changes 
on the attractor. One such tool is the characterization of the attractor through a complex network, 
which has gained increasing importance over the past one decade \cite {lac,mar1,yan,don1}. In this 
approach, the chaotic attractor is first transformed into a complex network using some suitable 
method. The resulting complex network, which is characteristic to the structure of the attractor, 
is then analysed using standard network measures \cite {str,new,bar}. This approach has, over the 
years, become a mature field with several new methods and ideas emerging, especially regarding 
the characterization of attractors re-constructed from time series.

In this work, we concentrate on a key measure, namely entropy, and show its effectiveness in quantifying 
structural information loss on chaotic attractors using measures of recurrence networks. In fact, network 
entropy has been proposed by various authors 
\cite {band,dem,saf,bia} in different contexts in the literature, whose details are given in the next section. 
However, none of them have addressed the specific issue of information loss on a chaotic attractor through 
network measures.  Here, we show that network based analysis is useful to quantify information 
loss associated with the change in the structure of chaotic attractors. For this, we propose a new 
\emph {entropy measure} suitable for the analysis of recurrence networks by modifying the basic equation for 
network entropy already existing in the literature. We also comment on some practical 
aspects of the proposed measure and its relevance in the context of the theory of complex networks. 

Our paper is organized as follows: A brief review of the existing network entropy measures is given in 
the next section where we also present the details of the entropy measure that we propose in this work with 
specific emphasis to quantify the information loss on chaotic attractors due to structural changes.  
The complex network that we construct from the time series 
is called the \emph{recurrence network}, since the specific property of recurrence \cite {eck} 
of dynamics is used for its construction. The details of this construction procedure are presented in 
\S 3, where we apply the proposed entropy measure for the 
analysis of recurrence networks from chaotic attractors. In \S 4, we show how the proposed measure can be 
used to quantify the information loss on chaotic attractors in terms of the changes in the link density of 
the corresponding recurrence networks. In \S 5, we show that the measure proposed here is useful in the 
analysis of real world data as well. The paper is concluded in \S 6. 

\section{Entropy measure for complex networks}
Entropy is a key concept originating from statistical thermodynamics \cite {bal} and 
effectively used in information theory \cite {cov} and in the theory of dynamical 
systems \cite {bec}. In the realm of complex networks, entropy has been defined in different 
contexts, to measure the robustness and stability of a network \cite {dem}, as a measure of 
complexity \cite {band,bia} and to predict the most probable network under the given constraints 
using the maximum entropy principle \cite {saf}. There are competing notions of graph entropy  
measures applied in such diverse fields as computer science, information theory, sociology, 
chemistry and biology. What may be useful in one domain may not be in another. For example, 
graph entropy has been used to characterize the structure of graph based systems in mathematical 
chemistry \cite {bon}, where it serves as a complexity measure. Details regarding the history of 
graph entropy measures can be found in the review by Dehmer and Mowshowitz \cite {deh}. 

In this work, we consider an entropy measure very close to the \emph{structural entropy} of 
network ensembles as given by Karthik and Bianconi \cite {ana}. The authors, however, use a 
statistical mechanical approach where entropy of a network ensemble is proportional to 
$\log$ number of networks in the ensemble. By defining free energy on the ensemble, they show 
that scale free degree distributions have small value of structural entropy 
while Poisson degree distributions are more likely with large structural entropy.

Here we define an entropy measure in terms of the uncertainty in information content on a complex 
network with links as channels of information transfer between the nodes. Note that the term 
\emph{information} is used here in a general sense depending on what system or process the 
complex network represents. Consider an un-weighted and un-directed complex network of any 
topology of $N$ nodes and let the $\imath^{th}$ node has degree $k_i$, where $1 \leq k_i \leq (N-1)$, 
with no isolated node present in the network. 
Note that $\imath$ is an index representing the node number here and many nodes can share the 
same value of degree. We now define the probability of information transfer through the 
$\imath^{th}$ node as: 
\begin{equation}
 p_i = {{k_i} \over {(N-1)}}
 \label{eq:1}
\end{equation}
since the node is connected to only $k_i$ other nodes out of a maximum possible 
$(N-1)$ nodes.

We now introduce a new measure analogous to the structural entropy of network as: 
 
\begin{equation}
 E_m = {{-1} \over {N}} [\sum_{i}^N p_i \log p_i + \sum_{i}^N (1-p_i) \log (1-p_i)]
 \label{eq:2}
\end{equation}

Note that the first term in the above equation is the Shannon entropy of degree distribution which 
has been discussed in various contexts \cite {zou}. The novelty here is to add the second term 
and it implies that the measure is decided not only by 
the presence of link between two nodes, but also by its absence denoted by the 
remaining degree \cite {sol}.  
We show below that the proposed measure can characterize the uncertainty of information 
associated with the structural changes of a complex network. Hence we call it the 
entropy measure of a network. 

If all $k_i = 0$, then there is no information transfer and $E_m = 0$. If all the nodes are 
mutually connected, that is, $k_i = <k> = (N-1)$, then the information content becomes saturated 
and no new information can be generated from the network (or the function of the network 
disappears) and $E_m = \log 1 = 0$. Suppose we now consider any regular or homogeneous network 
with all $k_i = <k>$ less than $(N-1)$, then we can show that 
\begin{equation}
 E_m^h = -[p \log p - p \log (1-p) + \log (1-p)]
 \label{eq:3}
\end{equation}
where $p = {{<k>} \over {(N-1)}}$. Note that the superscript $h$ in the above equation is used to 
denote homogeneous networks. Thus, for an un-weighted and un-directed complex network of $N$ nodes 
with all $k_i = <k>$, $E_m^h$ first increases as $<k>$ increases from zero, becomes a maximum and 
decreases to zero again as $<k> \rightarrow (N-1)$. In other words, $E_m^h$ varies as a one-hump 
function with $<k>$.

We now represent the above equation for $E_m^h$ in terms of the link density of the network. Any 
complex network, in general, has a range of values for $k_i$, from $k_{min}$ to $k_{max}$. The 
total number of links in the network is given by:
\begin{equation}
 L = {{1} \over {2}} \sum_{i=1}^N k_i
 \label{eq:4}
\end{equation}
since a link is shared by two nodes. If we assume a \emph{homogeneous} version of this network 
(with $L$ fixed), the degree of each node is:
\begin{equation}
 <k> = {{\sum_i k_i} \over {N}} = {{2L} \over {N}}
 \label{eq:5}
\end{equation} 
Therefore we get the probability associated with each node is:
\begin{equation}
 p = {{<k>} \over {(N-1)}} = {{2L} \over {N (N-1)}} = \rho
 \label{eq:6}
\end{equation}
where $\rho$ is the link density of the network. From above, the entropy measure of this network is 
given by:
\begin{equation}
 E_m^h (\rho) = -[\rho \log \rho - \rho \log (1-\rho) + \log (1-\rho)]
 \label{eq:7}
\end{equation}
As $\rho \rightarrow 0$ $(<k>=0)$ and $\rho \rightarrow 1$ $(<k>=(N-1))$, the entropy measure 
tends to zero. In between, at an optimum value of $\rho$, $E_m$ becomes maximum. 

Note that, we have an ensemble of networks with  $\rho$ fixed (for any $N$) and each element of this 
ensemble has a different degree distribution and different value of $E_m$. Out of this ensemble, 
there is one element (since all nodes are equivalent) which is homogeneous having the maximum value 
of entropy measure, $E_m^h|_{\rho}$. Entropy of all other elements is less than that of the 
homogeneous network: $E_m|_{\rho} < E_m^h|_{\rho}$. This also implies that the entropy measure 
proposed here is minimum for the completely heterogeneous network \cite {rj4}, with each $k$ 
value different from one another: $\{k\} \equiv \{1,2,3,....(N-1)\}$. Thus, for an ensemble of 
networks with given $N$ and $\rho$, $E_m$ increases from a minimum value $E_m^{het}$ to a maximum value 
$E_m^h$. 

The above result can be understood from an information content point of view as follows: 
Entropy is a measure of uncertainty in a system or process and hence $E_m$ can be considered as 
a measure of uncertainty in information content associated with a network. When the network 
characterizing a system or process becomes completely homogeneous, the \emph{information loss} 
is maximum, equal to $E_m^h$ and vice versa. In other words, $|E_m^h(\rho) - E_m(\rho)|$ can be taken 
as a measure of the \emph{information to be generated} for transforming a homogeneous network to any 
other complex network of arbitrary degree distribution of the same link density $\rho$. It should be 
noted that this result is independent of $N$ due to the normalization of the measure with respect to 
$N$.

We now try to quantify the above information loss in terms of the link density by relaxing the 
condition for constant $\rho$. This implies changes in the entropy measure due to 
a change in the number of links in the network rather than their re-distribution between the nodes. 
We have shown that the variation of $E_m^h$ for homogeneous networks 
as a function of $\rho$ has the form of a one-hump curve. So there is an optimum value, say 
$\rho_{opt}$, at which $E_m^h$ is maximum. This value of $\rho$ can be obtained by using the condition:
\begin{equation}
 {{d} \over {d \rho}} E_m^h (\rho) = 0
 \label{eq:8}
\end{equation}   
Using Eq. (7) and simplifying, we get $\rho_{opt} = {{1} \over {2}}$. Putting this back in Eq. (7), 
the maximum value of entropy for any homogeneous network is $E_m^h|_{max} = \log 2$. Also, there are 
two values of $\rho$ that lead to the same value of $E_m$, since its value remains unchanged by 
replacing $p$ by $(1-p)$ in the defining equation. In other words, one can find a complementary 
network leading to the same value of $E_m^h$. We now show that the above results can be used for 
quantifying information loss on chaotic attractors through RNs, whose details are given in the 
next section.

\section{Recurrence networks and entropy measure}
The first step in the network based approach is to transform the time series to a complex 
network. Several schemes \cite {zha,xu,gao} have been proposed for this over the last two 
decades and each one is useful in particular contexts of application. Here we use a specific 
scheme based on the property of recurrence of every dynamical system.  The resulting network,   
called the recurrence network (RN) \cite {don2,don3}, captures the structure of the embedded attractor.  
The method of RN is useful for the analysis of both synthetic and real world data 
and has found numerous applications from identifying dynamical transitions \cite {dong} and 
extreme events \cite {jkr} to detecting epileptic states in biomedical time series \cite {mar2} 
and has opened up a new window in the pursuit of complexity in real world systems, such as, 
climate data analysis \cite {mar3,boe}.

To transform the time series 
into a RN, it is first embedded in a multi-variate state space of dimension $M$ using the time 
delay co-ordinates \cite {gra}. Every point on the reconstructed attractor is then identified as a node 
and a recurrence threshold ($\epsilon$) is set to define the connection between two nodes. Two nodes 
are considered to be connected if the corresponding points on the attractor are within the limit of 
this threshold. From the construction, it is clear that the RN is an un-weighted and un-directed 
network with the elements of the adjacency matrix $A_{\imath \jmath}$ either $1$ or $0$ depending on 
whether two nodes are connected or not.

The choice of the parameter $\epsilon$ is crucial for the construction of the RN to ensure that 
the resulting network is a proper representation of the embedded attractor. We have recently 
proposed a scheme \cite {rj1} for the construction of RN where the value of $\epsilon$ is automated 
for a given embedding dimension $M$. Here the time series is first transformed into a uniform 
deviate so that the size of the embedded attractor is rescaled into a unit $M$-cube. A critical 
range of threshold $\Delta \epsilon$ is selected for the construction of the network whose minimum 
is chosen with the standard condition that there exists a single giant component in the resulting 
RN. We have shown that \cite {rj1} this critical range is approximately identical for several 
chaotic systems and white noise and depends only on the embedding dimension $M$. The threshold 
values used here for the construction of the network are $\epsilon = 0.06$, $0.10$, $0.14$ and 
$0.18$ for $M$ varying from $2$ to $5$ respectively.   
The scheme has been effectively employed to study the influence 
of noise on the structure of chaotic attractors \cite {rj2} and for the analysis of light curves 
from a prominent black hole system \cite {rj3}. Here we use this scheme for the construction of 
RN from time series.  

All the results presented in \S 2 are true for RNs as well, as they are un-weighted and un-directed 
complex networks. We first show that the proposed entropy measure is a characteristic property 
of every chaotic attractor, independent of the parameters used for the construction of the network. 
For this, we construct RNs from time series of several standard chaotic attractors and white noise 
using the scheme mentioned above \cite {rj1}. The RNs are constructed for differerent fixed values of 
$N$ by changing the embedding dimension $M$ and vice versa. 

\begin{figure}
\begin{center}
\includegraphics*[width=16cm]{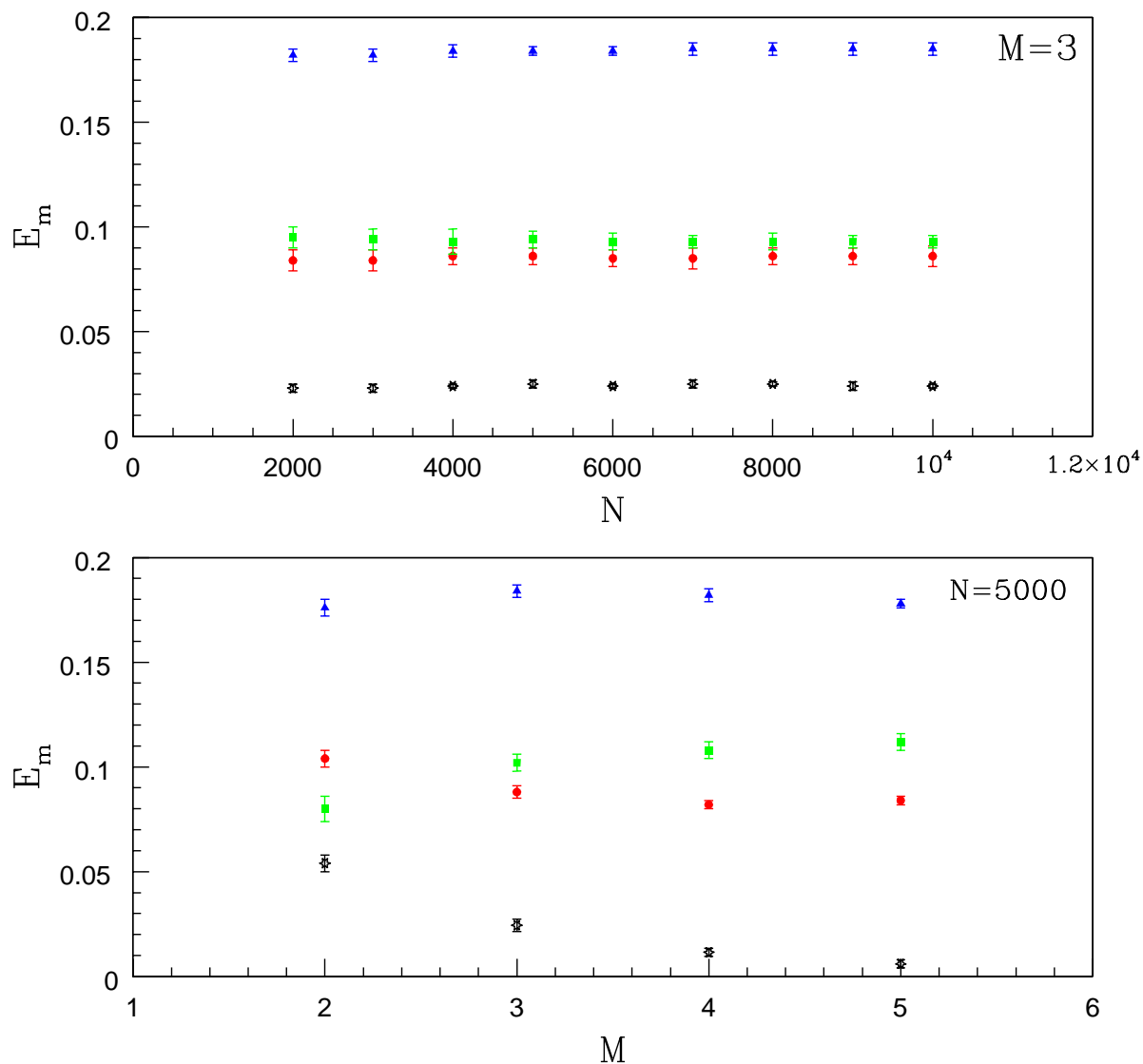}
\end{center}
\caption{Evidence to show that the proposed entropy measure is a characteristic property 
of every chaotic attractor. Results for three standard chaotic attractors are shown along with 
that for white noise. Top panel shows the variation of the measure with respect to the 
number of nodes $N$ in the network for a given embedding dimension $M$ and bottom panel 
vice versa. In both cases, solid circles are values for the RN from Lorenz attractor, solid 
squares for R\"ossler attractor, solid triangles for Henon attractor and asterisk for 
white noise.} 
\label{f.1}
\end{figure}

The variation of $E_m$ for RNs from three standard chaotic attractors and white noise as a function of 
$N$ and $M$ are shown in Fig.~\ref{f.1}. While the values saturate for all the chaotic attractors with 
respect to both $N$ and $M$, it shows a decreasing trend for white noise as a function of $M$ 
(bottom panel). The calculations are repeated for $10$ different time series in each case and the 
average values are shown in the figure with the standard deviation as the error bar. It is clear that 
the proposed measure is a characteristic property of every chaotic attractor independent of both 
$N$ and $M$.

\begin{figure}
\begin{center}
\includegraphics*[width=16cm]{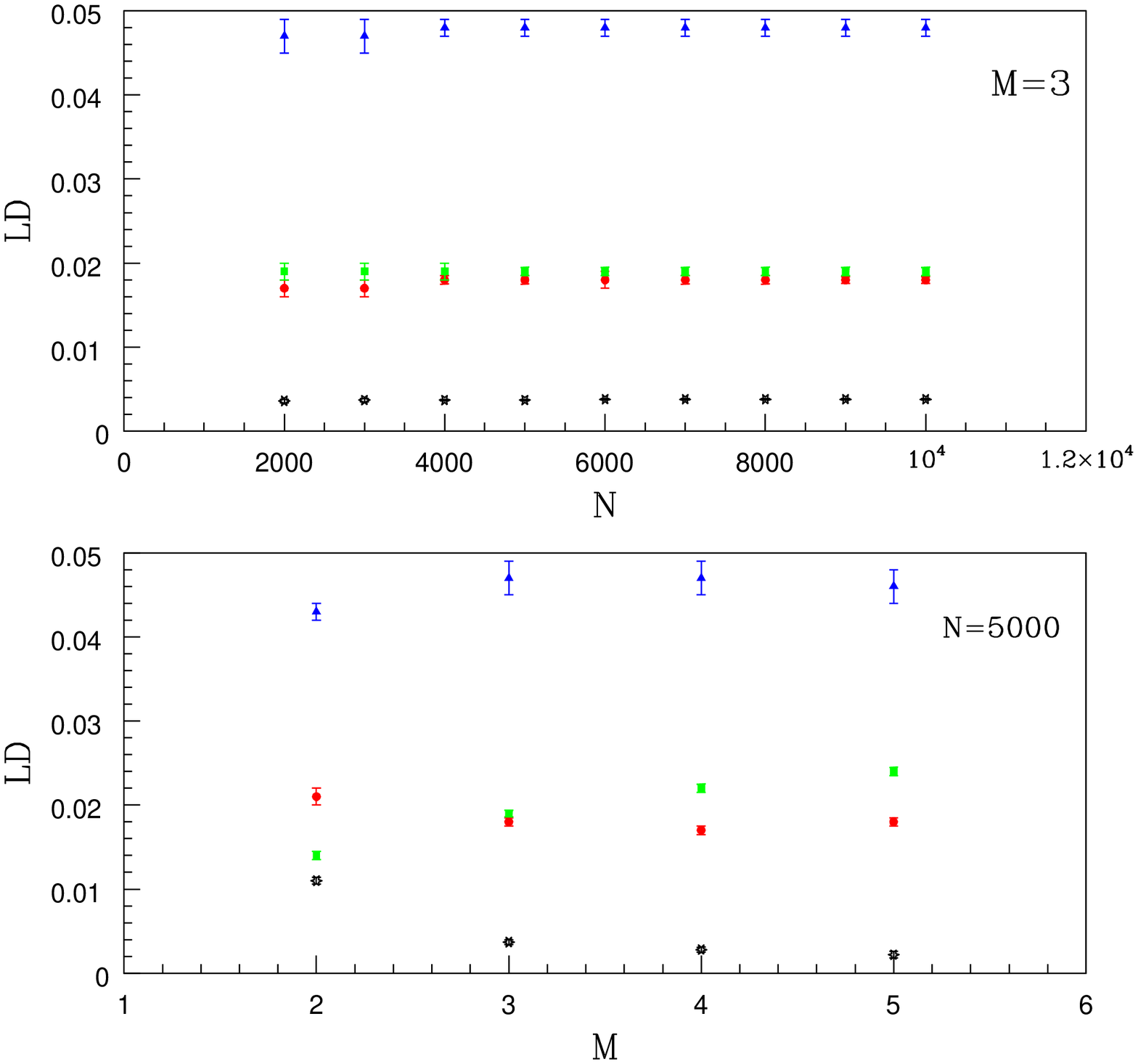}
\end{center}
\caption{Variations in the link densities as a function of $N$ and $M$ for the same RNs whose 
entropies are plotted in the previous figure. The notations of systems are the same as in the 
previous figure.} 
\label{f.2}
\end{figure}

Since the proposed measure is closely related to the link density of the network as shown in the previous 
section, we also check how the link density (LD) varies with $N$ and $M$. These results are shown in 
Fig.~\ref{f.2}, where variation in link density is shown with $N$ and $M$ for the same networks used in 
the previous figure. It is evident that the behaviour is identical to that of $E_m$. Also, the values of the 
LD for a given $N$ and $M$ are widely different between Henon, Lorenz and white noise. This is due to the 
difference in the structure of the RN which, in turn, depends on the structure of the embedded attractor. 
This implies that one has to be careful in choosing a specific value of LD commonly for all systems as a 
criterion for selecting the recurrence threshold for constructing the network.   

\begin{figure}
\begin{center}
\includegraphics*[width=16cm]{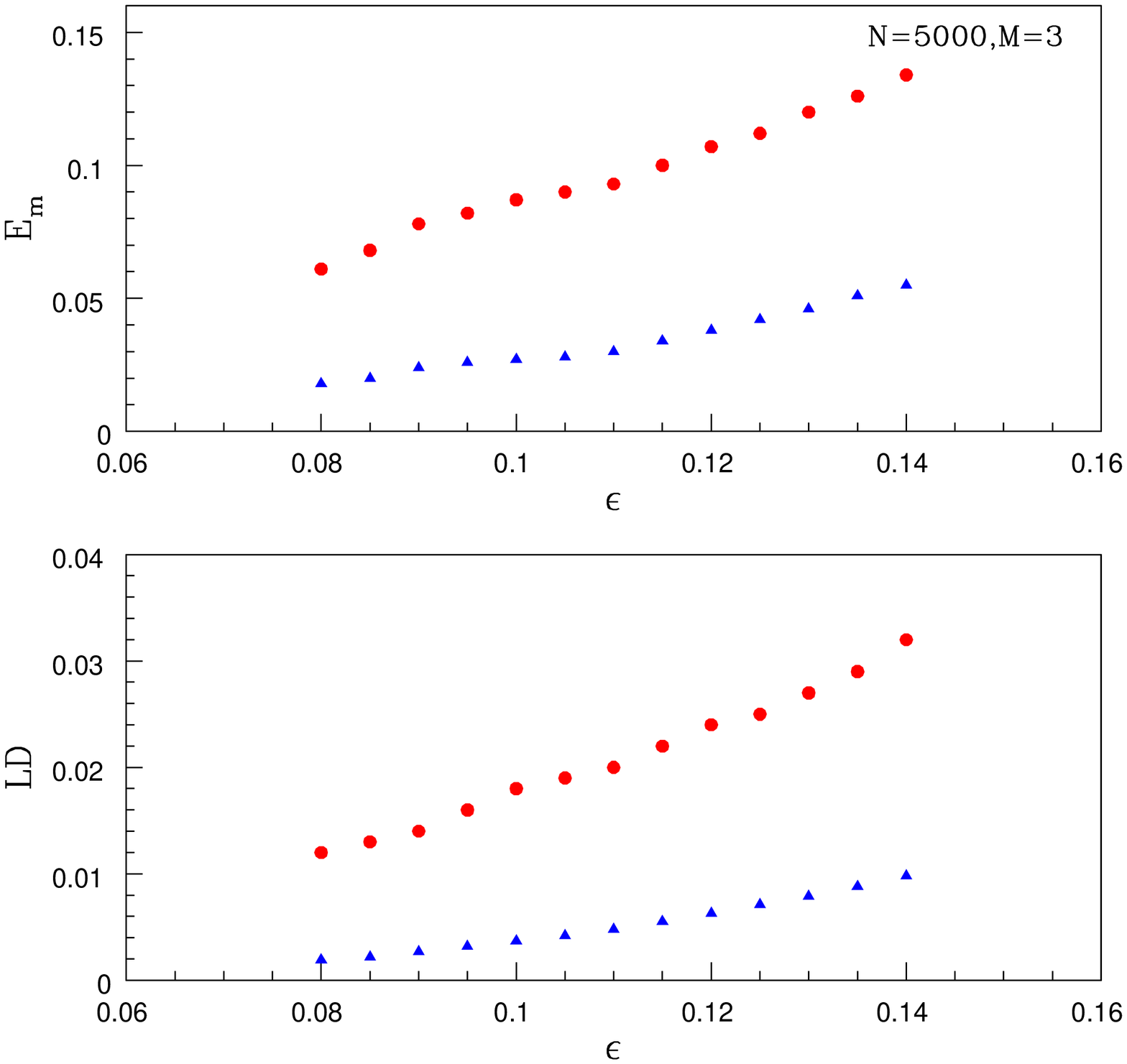}
\end{center}
\caption{Variation of $E_m$ with $\epsilon$ for RN from Lorenz attractor time series (solid circles) 
and white noise (solid triangles) is shown in top panel. The corresponding variation in LD is shown in 
the bottom panel.} 
\label{f.3}
\end{figure}

Here we have used a specific scheme for the choice of the recurrence threshold $\epsilon$ and the construction 
of the RN from time series which is based on our previous results. As is well known, there is no 
standard procedure for the choice of $\epsilon$ and there are, in fact, several suggestions for choosing it 
in the literature pertaining to different contexts and applications. A more detailed discussion regarding this 
can be found in the very recent review by Zou et al. \cite {zou}. Hence it is important to know how the 
measure proposed here is affected by small changes in $\epsilon$. This is shown in Fig.~\ref{f.3} 
(top panel). Here we show the variation of $E_m$ with $\epsilon$ for the RNs constructed from Lorenz 
attractor time series and white noise. In the bottom panel of the same figure, we show the corresponding 
variation in LD with $\epsilon$. As $\epsilon$ increases, it is natural that the LD increases due to 
increase in connections. This change is also reflected in $E_m$. However, around the threshold value used 
for the construction of the networks, this change is comparatively small in both cases vindicating our 
choice of $\epsilon$.   

\begin{figure}
\begin{center}
\includegraphics*[width=16cm]{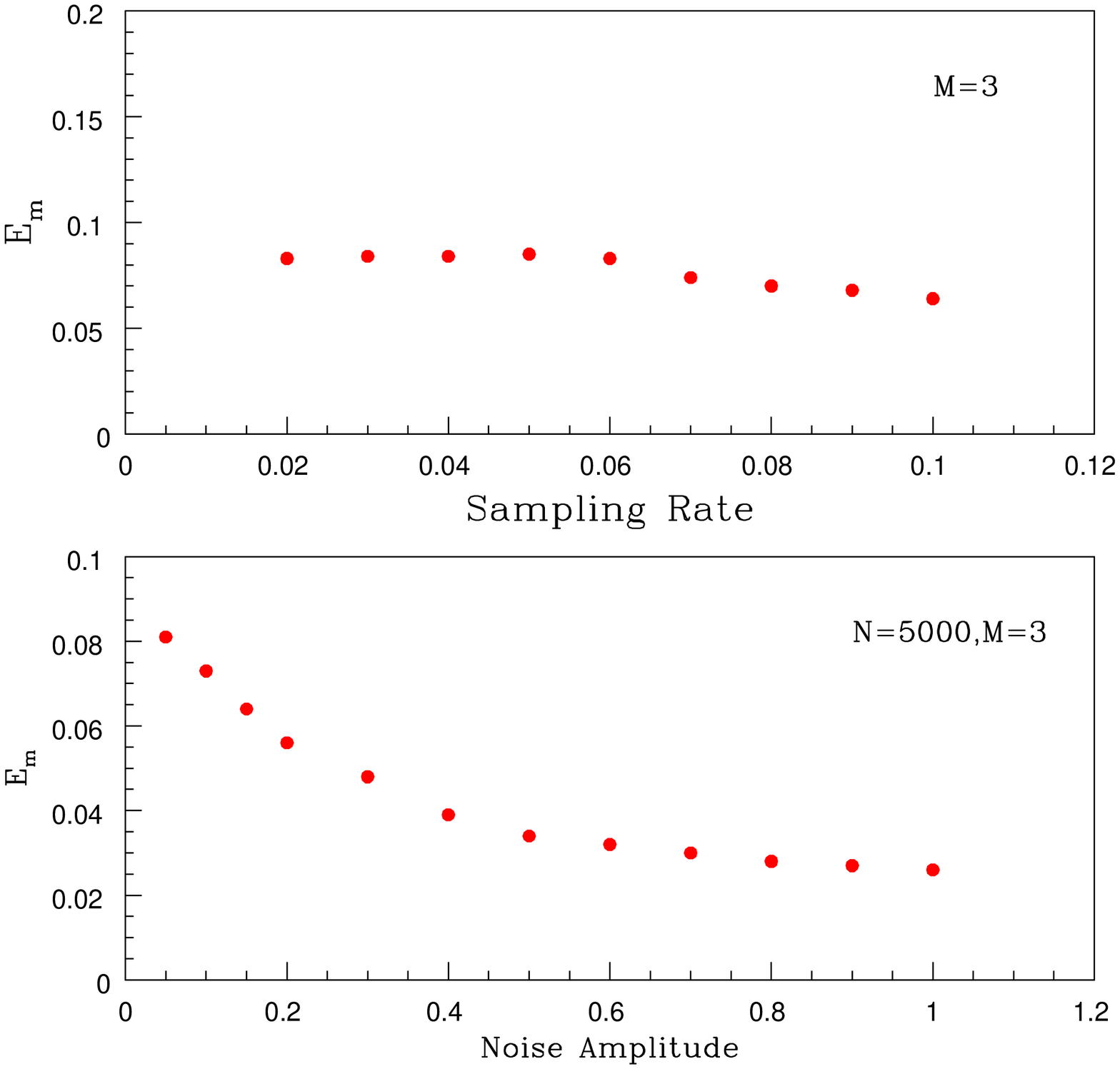}
\end{center}
\caption{The effect of sampling rate $\Delta t$ used to generate the Lorenz attractor time series on the 
proposed measure is shown in the top panel. As $\Delta t \rightarrow 0.1$, there is a decrease in $E_m$ indicating 
oversampling. The change in $E_m$ for the addition of different percentages of white noise on Lorenz 
attractor time series is shown in the bottom panel.} 
\label{f.4}
\end{figure}

It is also important to check the dependence of $E_m$ on other parameters involved in the construction of 
RN. We have already shown in Fig.~\ref{f.1} that the measure is stable with respect to changes in $N$ 
and $M$ as per our scheme. We now check the dependence of $E_m$ on the sampling rate $\Delta t$ for the 
generation of chaotic time series. This shown in Fig.~\ref{f.4} (top panel). We show the results for 
the variation of $E_m$ for the Lorenz attractor generated with sampling rates increasing 
from $0.02$ to $0.1$. The total time of evolution in each case is fixed as $200$ seconds (increasing the length of 
time series as sampling rate decreases) to ensure there are enough trajectory points on the attractor. 
The values of $E_m$ are approximately constant for the first few sampling rates, but decreases as  
$\Delta t \rightarrow 0.1$.  
This is an indication that as $\Delta t$ becomes large, the attractor gets slightly over sampled and the 
finer details of the 
Lorenz attractor is lost. We have used $\Delta t =0.05$ in our computations in this work.

Finally, in the analysis of real world data, noise contamination is always present. We now check how 
$E_m$ is affected by additive noise in a time series. A more detailed discussion on this is given in 
\S 4. We construct RNs from Lorenz attractor time series by adding different percentages of white 
noise from $5\%$ to $100\%$. Results of computation of $E_m$ for these RNs are also shown in 
Fig.~\ref{f.4} (bottom panel). It is clear that as the noise amplitude increases, there is a 
systematic decrease in the value of $E_m$.      
    
\section{Entropy measure and information loss}
Let us now consider the RN constructed from a typical chaotic attractor for a fixed $N$. The topology and the 
structure of the network will be characteristic to that of the attractor \cite {rj2}, with a range of values  
for ${k_i}$, a characteristic degree distribution and an entropy measure denoted by $E_m^c$. 
Imagine an ensemble of networks with the same link density $\rho_c$ of which the RN is just one element. 

\begin{figure}
\begin{center}
\includegraphics*[width=16cm]{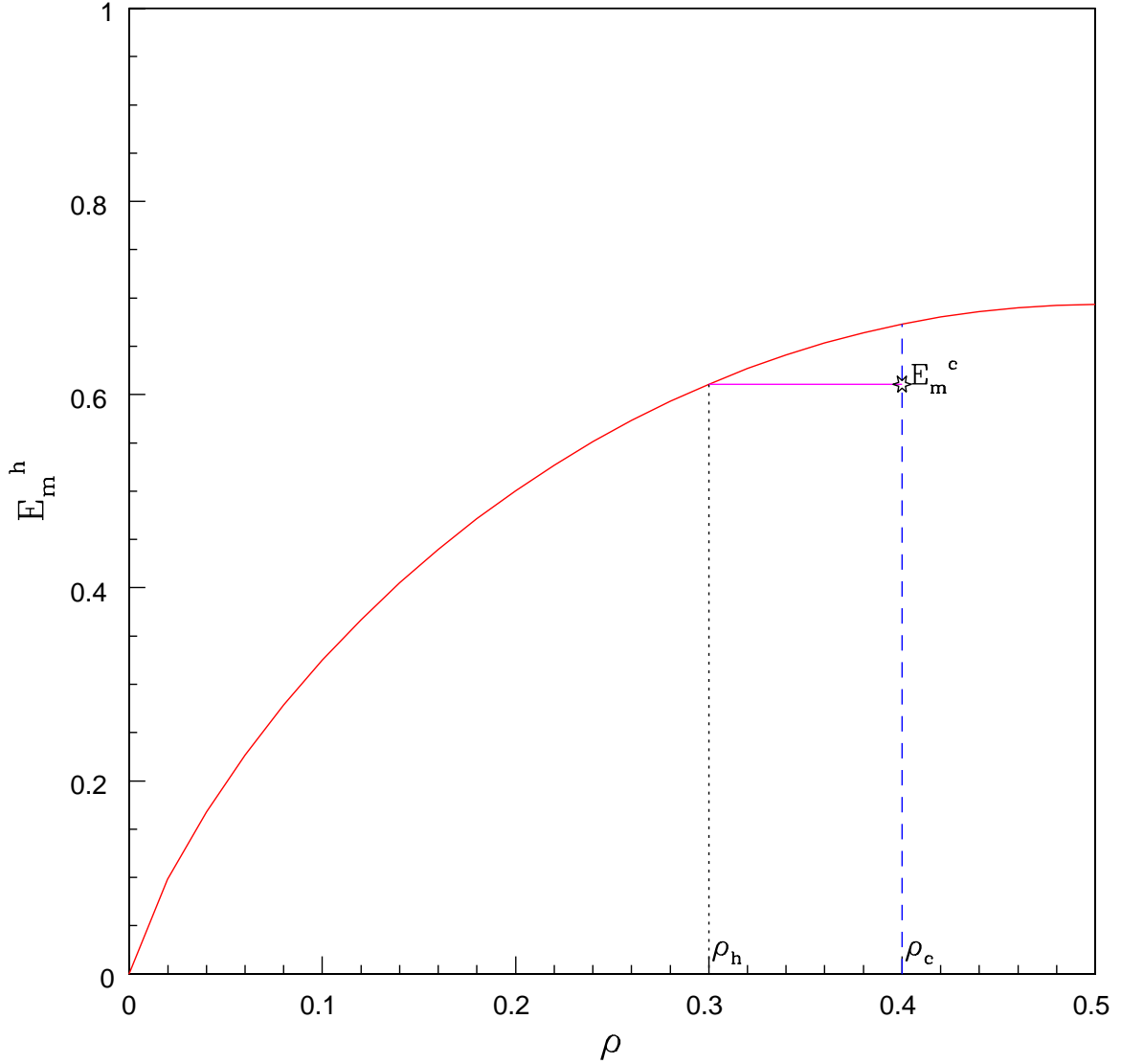}
\end{center}
\caption{The solid curve represents the variation of the entropy measure $E_m^h$ of the 
homogeneous networks of any $N$ as the link density $\rho$ varies from $0$ to $0.5$. 
The variation for $\rho \in [0.5,1]$ is a mirror reflection of the first half due to the 
symmetry of the measure with respect to exchange of $p$ and $(1-p)$ in the proposed 
equation. In the figure, $\rho_c$ represents the link density of the RN from a typical 
chaotic attractor whose entropy measure is denoted by the asterisk on the dashed vertical 
line. The vertical line correspond to an ensemble of networks with fixed link density 
$\rho_c$ differing in their topology or degree distribution, with the homogeneous element 
having the maximum value of the measure $E_m^h$. There is a homogeneous network whose 
entropy measure is same as that of the RN (denoted by the horizontal solid line), but with 
a reduced link density $\rho_h$. The difference $|\rho_c - \rho_h|$ can be taken as 
a measure of the information to be generated to obtain a specific RN from a trivial 
homogeneous network.} 
\label{f.5}
\end{figure}

From our discussion in \S 2, the element of this ensemble with the maximum entropy $(E_m^h)$ is the 
homogeneous network. In Fig.~\ref{f.5}, we show the variation of $E_m^h$ as $\rho$ varies from $0$ to $0.5$. 
The vertical dashed line in the figure represents the ensemble of networks with a fixed link density 
$\rho_c$ as that of the RN. The element corresponding to the point where the dashed line meets the curve 
is the homogeneous network with the maximum entropy for the fixed value of $\rho_c$. The point, represented 
by asterisk on the dashed line, corresponds to the RN from the attractor with the entropy measure denoted by 
$E_m^c$. Obviously, any RN can be obtained starting from a homogeneous network of same number of nodes and 
by \emph{re-distributing} the links (without changing the link density). By doing this, the entropy measure 
drops from $E_m^h$ to $E_m^c$. In other words, we can consider $|E_m^h - E_m^c| \equiv \Delta E_m$ 
as a measure of the information loss when a RN gets transformed into a homogeneous network of same $\rho$. 
Since $E_m$ is a characteristic property of the attractor, $\Delta E_m$ should also be characteristic 
to the attractor.

The value of $\Delta E_m$ also indicates how much the structure of a given network differs from a 
corresponding homogeneous 
network for a \emph{given link density}. For example, the value of $E_m$ of the RN from white noise is 
found to be much 
less compared to that of the RN from a chaotic attractor. However, the value of $E_m^h$, which depends on the 
link density, is also very small for the homogeneous network corresponding to 
white noise making its $\Delta E_m$  smaller compared to 
that of the RN from any chaotic attractor. 

\begin{table}[h]
\centering
\begin{tabular}{|l|c|c|c|c|}
\hline
\emph{System} & $\Delta E_m$ & $\rho_c$ & $\rho_h$ &  $\Delta \rho$  \\
\hline
\hline

Lorenz &  0.0034 & 0.0184 & 0.0162  & 0.0022  \\

& &  &  & \\

R\"ossler & 0.0018 & 0.0210 & 0.0194  & 0.0016  \\

& &  &  & \\

Henon & 0.0037 & 0.0448 & 0.0421  & 0.0027  \\

& &  &  & \\

White Noise & 0.0004 & 0.0037 & 0.0036  & 0.0001 \\

\hline
\hline
\end{tabular}
\caption{Difference in link density $\Delta \rho$ as an index of information loss for some standard 
chaotic attractors and white noise, along with the corresponding change in the entropy measure 
$\Delta E_m$.}
\label{tab:1}
\end{table}

We now try to represent the above information loss in terms of the link density, since it is a directly 
measurable parameter for any complex network. For this, again consider 
Fig.~\ref{f.5}. There is a homogeneous network with the same value of entropy as that of the RN as 
indicated by the horizontal solid line. This homogeneous network has a reduced link density $\rho_h$ 
denoted by the vertical dotted line. Suppose the given RN is transformed into a homogeneous network 
with the reduced link density $\rho_h$. Then, from the figure, its entropy remains unchanged at $E_m^c$. 
In other words, the loss of information can now be obtained in terms of a measurable change 
$\Delta \rho = |\rho_c - \rho_h|$ of two homogeneous networks, which varies directly with the 
corresponding change in entropy $\Delta E_m$.  In Table I, we show the values of $\Delta E_m$ and 
$\Delta \rho$ for the RNs from some standard chaotic 
attractors and white noise. Note that, the RN from white noise is very close to a 
homogeneous network with the degree of each node centered around an average value $<k>$ and hence 
both $\Delta E_m$ and $\Delta \rho \rightarrow 0$. 
Interestingly, the link density and the value of $E_m$ are 
slightly higher for the RN from the R\"ossler attractor compared to that of the Lorenz attractor. However, 
$\Delta E_m$ and $\Delta \rho$ are higher for the latter, indicating that its RN is more 
different from a homogeneous network compared to that of R\"ossler.  
It is thus clear that the proposed measure is 
suitable for characterizing the structural changes of a chaotic attractor. 

\begin{figure}
\begin{center}
\includegraphics*[width=16cm]{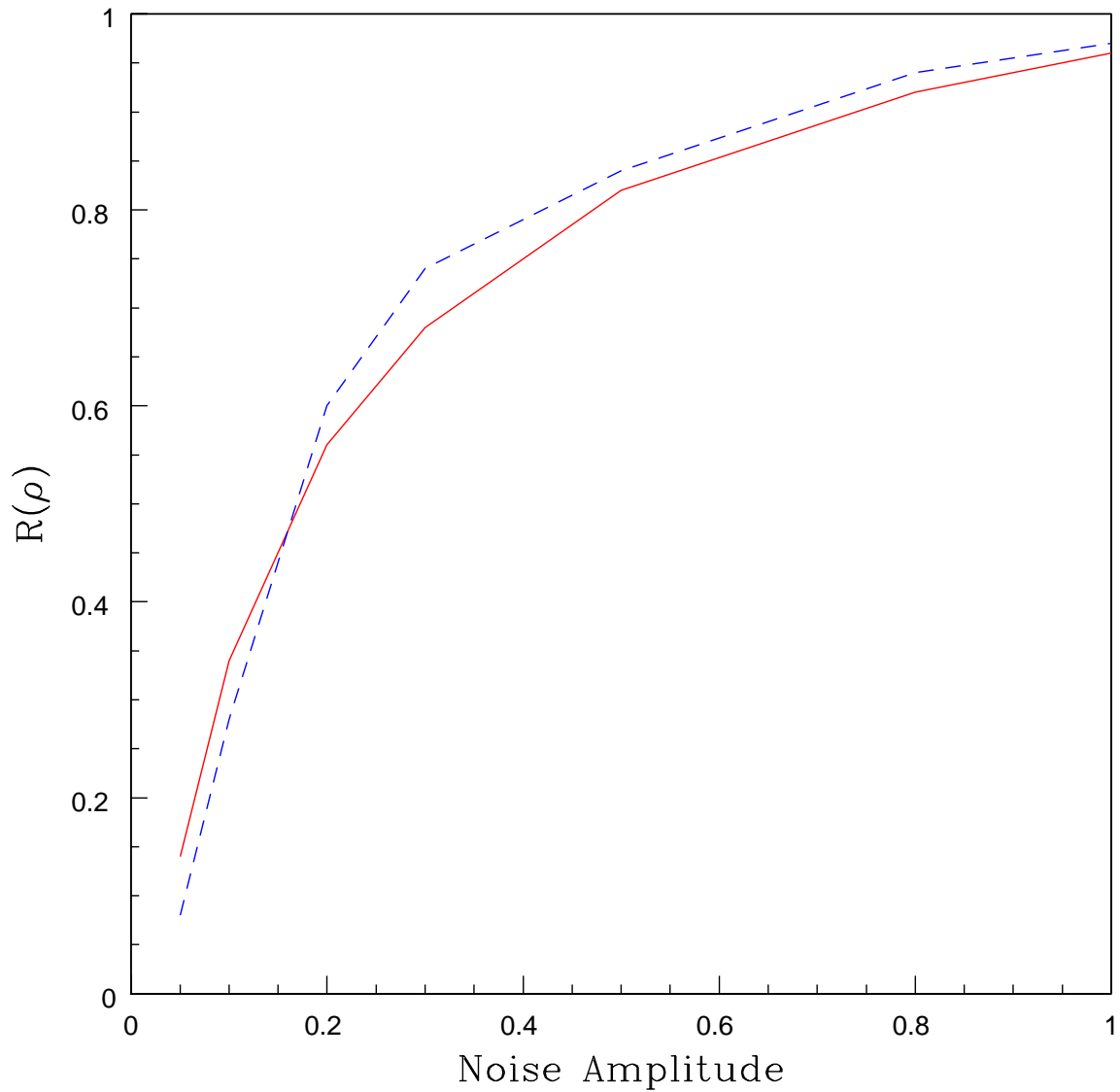}
\end{center}
\caption{Variation of the ratio of link densities $R(\rho)$ (see text) as a function of noise 
amplitude for the RNs from the Lorenz attractor (solid line) and the Henon attractor (dashed line) 
added with different percentage of noise. The variation is approximately identical for the two 
cases and saturate as the noise amplitude tends to 1.} 
\label{f.6}
\end{figure}

We now consider one such example of structural loss for a chaotic attractor, namely, due to 
contamination by noise. We have already shown \cite {rj2} that when noise is added to a chaotic time 
series, the range of $k$ values in the constructed RN gets depleted. When the percentage of noise is 
sufficiently high, the structure of the attractor and hence the corresponding RN tends to that from a 
white noise. The noise affects the highly clustered nodes in the RN. As a consequence, the degree of 
each node tends to be approximately identical to some average value $<k>$, reducing the link density, 
and consequently the value of $E_m$. As the noise level increases, the difference in the value of 
$\rho$ with respect to that of the original RN (that is, $(\rho_c - \rho)$) also increases 
correspondingly. Since $\rho_c$ is characteristic to the attractor, the variation in the ratio 
${(\rho_c - \rho)} \over {(\rho_c - \rho_w)}$, which we denote by $R(\rho)$, with noise level can 
serve as a quantifier to study how the structure of the attractor is affected by noise. Here the 
denominator is a constant for a given attractor, with $\rho_w$ representing the link density of 
the RN from a pure white noise. In the absence of noise, $\rho = \rho_c$ and $R(\rho) = 0$. As the 
network becomes identical to the RN from white noise, $\rho = \rho_w$ and $R(\rho) = 1$. 
In Fig.~\ref{f.6}, we plot the variation of $R(\rho)$ for the RNs from two standard chaotic 
attractors as a function of noise amplitude. Note that, though the structure of the two attractors 
are very different, the variation of $R(\rho)$ is approximately identical. 

\section{Analysis of real world data}
In this section we consider whether the proposed measure is suitable for the analysis of real world data. 
To show this, we analyse light curves from a variable star. We choose a sample light curve, namely 
KIC 4484128, from the group of RRLyrae stars in Kepler archive where nonlinear behavior is reported in 
previous analysis \cite {lind,geor}. Continuous data segment without gaps sampled at $0.0204$ days involving 
$32000$ data points is generated and more details regarding the data can be found elsewhere \cite {geor}. 

\begin{figure}
\begin{center}
\includegraphics*[width=16cm]{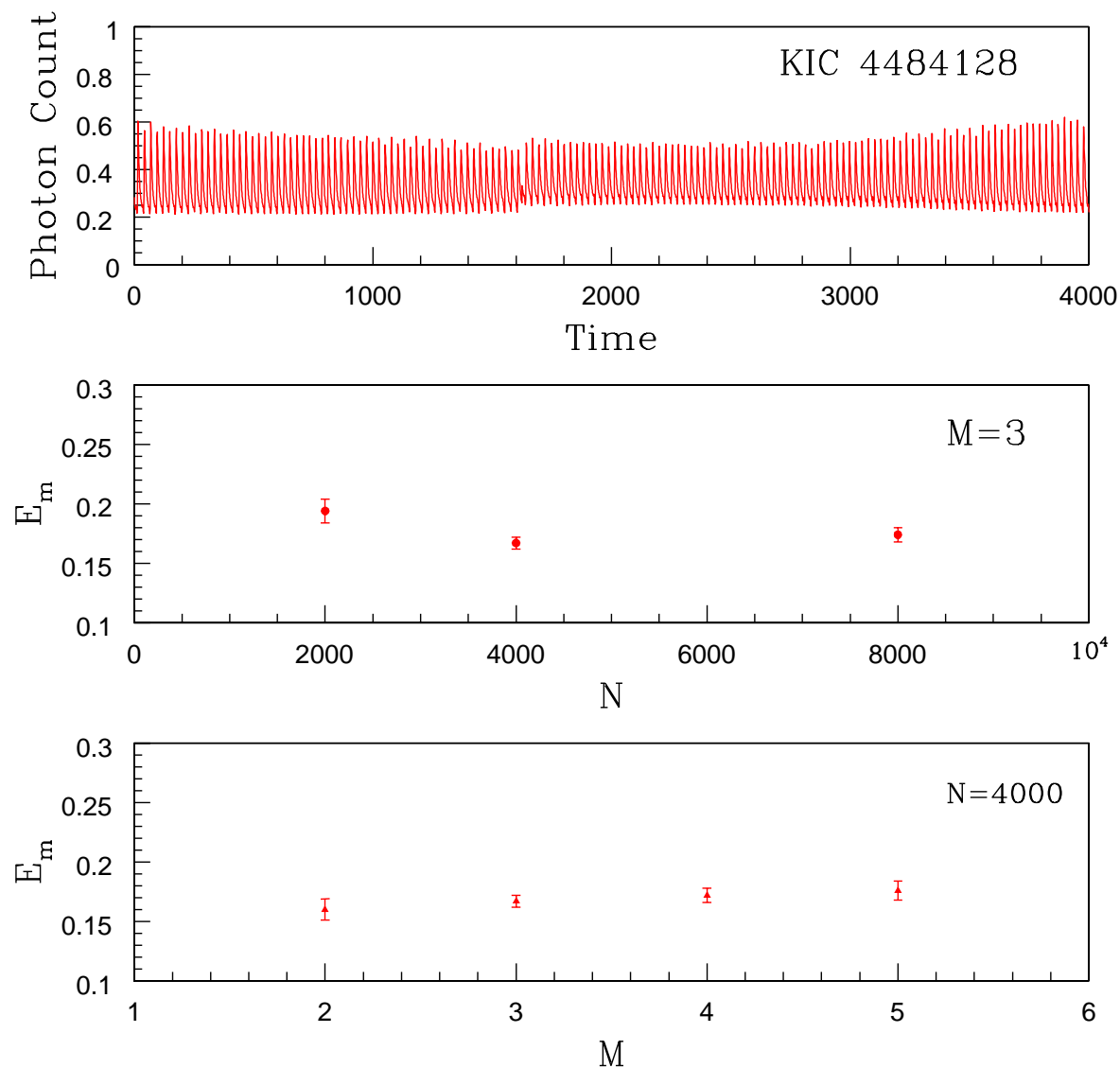}
\end{center}
\caption{A typical segment of light curve from the variable star KIC 4484128 from the Kepler archive is 
shown in top panel. The variation of $E_m$ for RNs constructed from the light curves for varying 
lengths with fixed $M$ and vice versa are shown in the middle and bottom panel respectively.} 
\label{f.7}
\end{figure}

For the present analysis, this full data is divided into different segments of varying lengths, of 
$2000$ data points (16 segments), $4000$ data points (8 segments) and $8000$ data points (4 segments). 
A typical segment is shown in Fig.~\ref{f.7} (top panel). RNs are constructed from each segment using 
our scheme and $E_m$ computed in each case. The results are shown in Fig.~\ref{f.7} (middle panel) for 
varying $N$ with $M$ fixed as 3 and vice versa in Fig.~\ref{f.7} (bottom panel) with $N$ fixed at 
$4000$. It is found that $E_m$ remains constant in both cases making it a robust measure useful for 
experimental time series. Moreover, we have also checked the $\Delta E_m$ and $\Delta \rho$ values for this 
data and find the average values of both as $\Delta E_m = 0.0066$ and $\Delta \rho = 0.0031$ for $N = 4000$ 
and $M = 3$, which is again an indication of inherent nonlinearity in the data. This, in turn, rules out 
pure stochastic process in the temporal variability of the light curve, as suspected earlier. 

Another possible practical application of the proposed measure in this context is the following: 
There are large number of variable stars in the Kepler Archive which have been classified from an 
astrophysical point of view based on their intensity and spectral variations. It is also possible to 
classify them from a dynamical perspective and noise content for a better understanding of their 
temporal variability. The measure proposed here seems to be suitable for this as it can capture 
subtle variations in the light curve as reflected in the RN. These measures are reliable even with a 
short time series of a few thousand data points as in the case of variable stars where long time series 
is difficult to get due to various practical difficulties of observation.    

\section{Conclusion}
Nonlinear time series analysis using complex network measures has become an important area of research 
over the last two decades. One method of constructing the network, based on the property of recurrence, 
has become very popular due to the large number of practical applications of the resulting RN. Eventhough 
several measures have been used for the recurrence network analysis of chaotic time series,  
the use of entropy has been limited in all the previous analysis. Specifically, none of them have 
addressed the issue of information loss on a chaotic attractor through network measures.

In this work, we propose an entropy measure for the recurrence network based analysis of chaotic time series. 
We show that the measure can uniquely characterize the structural information loss of a chaotic attractor 
in terms of the changes in the link density of the RN constructed from the attractor. 
We also indicate how the measure can be useful in practice in identifying the loss of information 
on the structure of a chaotic attractor due to noise contamination.

Another possible practical application of the proposed measure is to use it as a discriminating measure in 
the analysis of real data by looking at how much the network constructed from the data is different from a 
homogeneous network. Also, the method has specific advantage that the finiteness of the data will not 
affect the accuracy of the analysis since the measure is independent of $N$, unlike the dimension 
measures used for similar studies previously \cite {hil}. The time series measured or observed from a 
real world system or process can vary based on several factors, such as, inherent dynamical changes, 
parametric instability, the amount and nature of noise added, etc. One important application of 
nonlinear time series analysis is to discriminate the data by identifying the subtle changes using 
measures of nonlinear dynamics. We hope the proposed measure will also be useful in this regard. 

Finally, though our focus in this work is the information loss on chaotic attractors through RNs, the 
measure proposed here could also be useful in many other contexts in the formal theory of complex networks. 
One possible application which we indicate here is the study of diffusion process on complex networks 
\cite {gom}. In this context, the entropy represents the minimum amount of information necessary to describe 
the diffusion on the network and is sensitive to the network topology. A high entropy indicates a large 
randomness and easiness of propogation from one node to another. Thus, maximizing entropy rate allows us 
to extract optimal diffusion process providing tools for design of network structure with a wide range of 
practical applications.   

{\bf Acknowledgments}

KPH acknowledges the computing facilities in IUCAA, Pune.

\end{document}